\documentstyle[aas2pp4,psfig]{article}

\newcommand\mzon   {M$_{\odot}$}

\newcommand\micros  {$\mu$s}

\righthead{The broad-band power spectrum of SAX J1808.4--3658}
\slugcomment{Submitted to ApJ, July 1998}

\begin{document}

\title{The broad-band power spectrum of SAX J1808.4--3658}

\author{Rudy Wijnands \& Michiel van der Klis}

\affil{Astronomical Institute ``Anton Pannekoek'' and Center for High
Energy Astrophysics, University of Amsterdam, Kruislaan 403, NL-1098
SJ Amsterdam, The Netherlands; rudy@astro.uva.nl,
michiel@astro.uva.nl}

\begin{abstract}

We analyzed the rapid aperiodic X-ray variability of the recently
discovered millisecond X-ray pulsar SAX J1808.4--3658. The power
density spectrum is dominated by a strong band-limited noise
component, which follows a power-law with index 1.0--1.3 at high
frequencies with a break that varies between 0.25 and 1.6 Hz, below
which the spectrum is relatively flat.  Superimposed on this, a broad
bump is present with a centroid frequency that varies well correlated
with the break frequency between 2.4 and 12.0 Hz. These
characteristics are very similar to what is commonly seen in other
low-luminosity low-mass X-ray binaries.  Between 100 and 400 Hz a
third broad noise component is present similar to that recently
reported in the low-luminosity low-mass X-ray binary 4U 1728--34 (Ford
\& van der Klis 1998).  We find a similar high-frequency noise
component also in other low-luminosity neutron star systems.  We
conclude that at any given epoch the rapid aperiodic X-ray variability
of the millisecond X-ray pulsar is indistinguishable from that of
other low-luminosity neutron star systems.

However, contrary to what has been found in those sources the break
frequency of the band-limited noise does not have a strict correlation
with mass accretion rate. With decreasing mass accretion rate, the
break frequency first decreased, then increased again. This previously
unobserved behavior could be typical for the very low mass accretion
rates we observe near the end of the outburst of SAX J1808.4--3658, or
it could be related to it being a pulsar.

\end{abstract}


\section{Introduction \label{intro}}

Recently, using the {\it Rossi X-ray Timing Explorer} (RXTE), the
first (and so far only) accretion-powered millisecond X-ray pulsar was
discovered (Wijnands \& van der Klis 1998a; 1998b). This source is
positionally coincident with the known transient X-ray burster and
low-mass X-ray binary (LMXB) SAX J1808.4--3658 (in 't Zand et
al. 1998; Marshall 1998), making it the first source which exhibits
coherent X-ray pulsations and thermonuclear X-ray bursts.  The pulsar
is in an $\sim$2 hour binary with a companion star of less than 0.1
\mzon\, (Chakrabarty \& Morgan 1998a; 1998b).

This discovery verified the twenty-year old prediction that the
progenitors of the millisecond radio pulsars are LMXBs (see
Bhattacharya \& van den Heuvel 1991 for a review of millisecond radio
pulsars), for which a lot of circumstantial evidence was already
present (see Wijnands \& van der Klis 1998b for references).  The
derived magnetic field of the millisecond X-ray pulsar is in the range
of the fields of the millisecond radio pulsars (Wijnands \& van der
Klis 1998b; Chakrabarty \& Morgan 1998b; Gilfanov et al. 1998; Cui et
al. 1998), but it is lower than that estimated for some other LMXBs
using indirect methods (see Wijnands \& van der Klis 1998b for
references). It is therefore not clear why other LMXBs, particularly
those at similarly low accretion rates as SAX J1808.4--3658, are not
also millisecond pulsars. Perhaps previous field estimates are wrong
and the magnetic fields in the other LMXBs are lower than that in SAX
J1808.4--3658, or perhaps in SAX J1808.4--3658 some special geometric
condition, such as low inclination or large spin and magnetic axis
misalignment, uniquely allows it to pulse.

In this Letter we study the rapid aperiodic X-ray variability of this
millisecond X-ray pulsar and compare it with that of other
low-luminosity LMXBs.  We conclude that apart from the pulsations the
rapid variability of SAX J1808.4--3658 is very similar to that of
other low-luminosity neutron star LMXBs; we do, however, identify one
possible unusual aspect.

\section{Observations and Analysis  \label{observations}}

During the decay of its April 1998 outburst SAX J1808.4--3658 was
frequently observed with the RXTE proportional counter array (PCA)
until May 6 (Fig. \ref{parameters}a; see also Gilfanov et al. 1998;
Heindl \& Smith 1998; Cui et al. 1998). During these observations data
were collected with a time resolution of 122 \micros\, in 64 channels,
except on April 13th when a mode with 1 \micros\, in 256 channels was
used. We used these data to calculate 512 second FFTs. After
subtraction of the deadtime-modified Poisson level, the resulting
0.004--4096 Hz power density spectra were averaged to form 21
different spectra, one for each observation (although some
observations were combined in order to increase sensitivity). Each
average spectrum contains between 1 and 23 ksec of data. The spectra
were fitted with a fit function that is described in Section
\ref{results}.  The errors on the fit parameters were calculated using
$\Delta\chi^2=1.0$. The upper limits were calculated using
$\Delta\chi^2=2.71$, corresponding to 95\% confidence levels.

\section{Results \label{results}}

A typical power density spectrum of SAX J1808.4--3658 is shown in Fig.
\ref{comparison}a.  Besides the very significant periodic signal at
401 Hz (see Figure 1 of Wijnands \& van der Klis 1998b) which is
hardly visible in this broad-band representation, several noise
components are visible. The overall shape is dominated by a strong
broad band-limited noise component, which follows a rough power law at
high frequencies with a break near 1 Hz, below which the spectrum is
almost flat.  On top of this noise component, at frequencies near 10
Hz a bump is present, which may have some sub-structure.  At
frequencies above 100 Hz a third noise component is present which cuts
off around 200--300 Hz. This spectral shape is very similar to that
seen in other low-luminosity LMXBs; in Fig. \ref{comparison}b we show
for comparison a power spectrum of 4U 1728--34. We used archival
RXTE/PCA data of March 1, 1996 of 4U 1728--34 to calculate this power
spectrum (see also Ford \& van der Klis 1998).

We fitted the power spectra with a fit function consisting of a broken
power law (representing the band limited noise; $P\propto
\nu^{-\alpha_{\rm below, above}}$, with $\alpha_{\rm below, above}$
are the power law index below and above the break frequency,
respectively), a Lorentzian (representing the $\sim$10 Hz bump), and
an exponentially cut-off power law (representing the noise component
at higher frequency; $P\propto\nu^{-\beta}e^{-\nu/\nu_{\rm
cut-off}}$). The higher frequency noise component could also be fitted
with a broken power law or a broad Lorentzian; this did not affect its
measured rms amplitude.  Fig. \ref{powerspectra} illustrates the
changes in the power spectrum of SAX J1808.4--3658 during the course
of the outburst; changes in the break frequency are clearly seen.

The behavior of the fit parameters during the decay of the outburst
are plotted together with the 3--25 keV X-ray flux in
Fig. \ref{parameters}. Both the break frequency and the frequency of
the bump decreased (from 1.60 to 0.25 Hz and from 12 to 2.4 Hz,
respectively) while the flux dropped (Fig. \ref{parameters}c squares
and triangles). After a 5--7 day interval during which these
frequencies remained at their minimum values, they increased again (to
1.2 Hz and 5.7 Hz, respectively), while the flux went on decreasing.
After April 26 the flux dropped within less than 3 days by a factor of
20 (see also Gilfanov et al. 1998; Heindl \& Smith 1998; Cui et
al. 1998); at this time, the break frequency decreased again to about
0.6 Hz. By this time, the bump had become undetectable.

The rms amplitude (Fig. \ref{parameters}e squares; 2--60 keV;
integrated over 0.01-100 Hz) of the band-limited noise is
anti-correlated with its break frequency. The rms increased from 20\%
at the highest break frequencies to about 28\% at the lowest break
frequencies.  The power law index below the break frequency
($\alpha_{\rm below}$) increased slightly during the decay (from
$-0.1$ to $0.0$; Fig. \ref{parameters}b). The power law index above
the break ($\alpha_{\rm above}$) varied between 1.0 and 1.3
(Fig. \ref{parameters}b), with no clear correlation with X-ray flux or
the break frequency of the band-limited noise.  After 29 April no
broad band power was detectable any more. The 20--30\% rms amplitude
upper limits during these late stages of the outburst are not very
stringent and the presence of similar noise components as before can
not be excluded.

The rms amplitude of the $\sim$10 Hz bump gradually dropped from 10\%
to 4\% during the outburst (2--60 keV; Fig. \ref{parameters}e
triangles), without any correlation with its frequency. After April 23
it is no longer detectable with upper limits of 5--8 \%.  The ratio of
the FWHM and the frequency of this bump varied slightly between 0.5
and 1.0.

The third noise component, between 100 and 400 Hz, has an rms
amplitude (2--60 keV; integrated over 10--1000 Hz) between 13\% and
23\% (Fig. \ref{parameters}f).  The index $\beta$ of the cut-off power
law (Fig. \ref{parameters}d) increases from $\sim$--1.4 (thus the
noise component is peaked) on April 11 to $\sim$0.5 on April 20, then
decreased again to $\sim$--1.6 on April 25. During the last two
observations for which this noise component could be detected the
index was $\sim$0. The cut-off frequency and the rms amplitude vary in
good correlation to this and all are anti-correlated with the break
frequency of the band-limited noise component. These variations
together result in keeping the effective cut-off of this higher
frequency noise component approximately constant around 200 Hz.

We measured the rms amplitude as a function of the photon energy of
the above noise components on three occasions during the
decay. All components tend to increase slightly with photon
energy. However, this is not usually very significant.

No kHz QPOs are detected. The (conservative) upper limits for a QPO
with a FWHM of 150 Hz between 500 and 1500 Hz are between 7\% and 8\%
at the highest observed luminosities (April 11 and 13, respectively),
and 4\% or worse (depending on the source flux and the amount of
observing time) further along the decay of the outburst.  We also
searched for kHz QPOs using only the highest energy channels, but none
were found.

\section{Discussion \label{discussion}}

The rapid X-ray variability of SAX J1808.4--3658 is dominated by a
broad band-limited noise component with a break frequency between 0.25
and 1.6 Hz. During the initial decay of the outburst this frequency
was correlated with the mass accretion rate, but at lower luminosities
it was anti-correlated. On top of this noise component a broad bump is
detected whose frequency is strongly correlated with the break
frequency. The third component is a broad noise component between 100
and 400 Hz, whose properties also correlate to the break frequency. As
all characteristic frequencies describing the broad-band noise appear
to vary in correlation, it is possible that the three components
together describe one complex physical phenomenon rather than three
independent processes.

\begin{figure}[t]
\begin{center}
\begin{tabular}{c}
\psfig{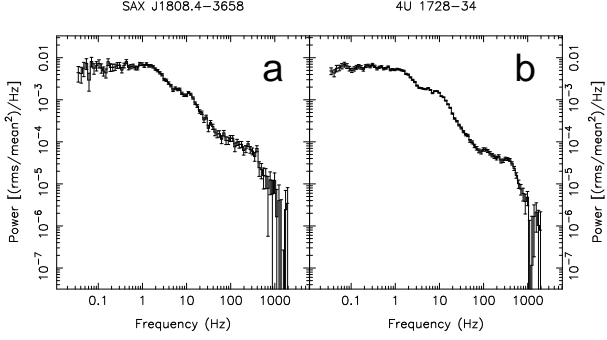}
\end{tabular}
\caption{Typical power spectra of SAX J1808.4--3658 ({\it a}) and 4U
1728--34 ({\it b}). The deadtime-modified Poisson level has been
subtracted. \label{comparison}}
\end{center}
\end{figure}

\begin{figure}[t]
\begin{center}
\begin{tabular}{c}
\psfig{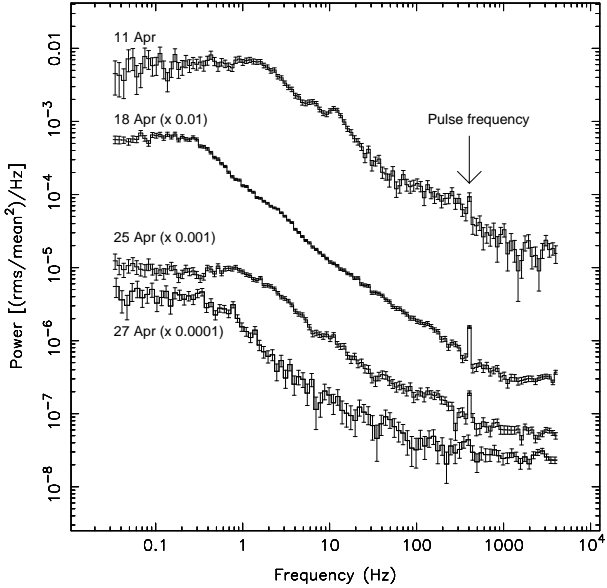}
\end{tabular}
\caption{Power density spectra of SAX J1808.4--3658 on different dates
during the 1998 outburst. The spectra appear flat at high frequencies,
as for display purposes only 97.5\% of the modified Poisson level was
subtracted.\label{powerspectra}}
\end{center}
\end{figure}

\begin{figure}[t]
\begin{center}
\begin{tabular}{c}
\psfig{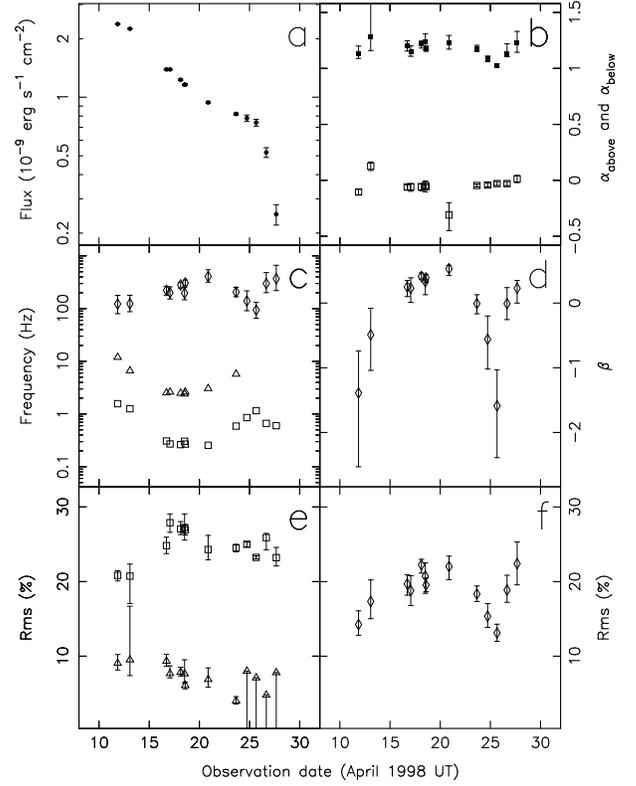}
\end{tabular}
\caption{The 3-25 keV flux ({\it a}; after Gilfanov et al. 1998), the
power law index of the band-limited noise above the break
frequency({\it b}; open squares; $\alpha_{\rm above}$), the power law
index of the band-limited noise below the break frequency ({\it b};
filled squares; $\alpha_{\rm below}$), the break frequency of the
band-limited noise ({\it c}; open squares), the frequency of the bump
({\it c}; open triangles), the cut-off frequency of the high frequency
noise component ({\it c}; open diamonds), the power law index of the
high frequency noise ({\it d}), the rms amplitude (2--60 keV)
of the band-limited noise ({\it e}; open squares), the rms amplitude
(2--60 keV) of the bump ({\it e}; open triangles), and the rms
amplitude (2--60 keV) of the high frequency noise component ({\it f})
during the decay of the 1998 outburst. Error bars are indicated except
where they are smaller than the data points.
\label{parameters}}
\end{center}
\end{figure}

Similar band-limited noise components are found in other LMXBs with
similar luminosities, such as 4U 1728--34 (Fig. \ref{comparison}b;
Ford \& van der Klis 1998).  The high frequency noise component was
not previously reported in other LMXBs than 4U 1728--34. We find by
re-examination of the data of 4U 0614+09 published by M\'endez et
al. (1997) that it is also present in this source. Also in the power
spectra published by Ford, van der Klis \& Kaaret (1998) of 4U
1705-44 a similar component can be seen.  It seems that this component
is a common feature in the power spectra of neutron star
low-luminosity LMXBs at low mass accretion rates, and not unique to
the millisecond X-ray pulsar.

So, the rapid X-ray variability below a few hundred Hertz between the
millisecond X-ray pulsar and low-luminosity neutron star LMXBs is
remarkably similar in terms of which components are present. There may
be differences in how these components depend on mass accretion
rate. From our detailed description of these dependencies in SAX
J1808.4--3658 one possible difference is already apparent: the break
frequency of SAX J1808.4--3658 does not have a strict correlation with
the mass accretion rate, whereas in the other sources it has been
observed that a strong positive correlation exists (e.g. Yoshida et
al. 1993; Prins \& van der Klis 1997; M\'endez \& van der Klis 1997;
M\'endez et al. 1997; Ford \& van der Klis 1998). However, an analysis
as detailed as we performed for SAX J1808.4--3658 has not yet been
done for when those sources were at their lowest mass accretion
rates. We can not exclude that there, also the break frequency is
anti-correlated with the mass accretion rates at certain (low)
luminosities.

We detect no kHz QPOs in SAX J1808.4--3658, with upper limits well
below detected amplitudes in other low-luminosity LMXBs (e.g. M\'endez
et al. 1997). However, in those sources below a critical luminosity no
kHz QPOs are detected either (M\'endez et al. 1997; Ford \& van der
Klis 1998).  In one case, 4U 0614+09 (M\'endez et al. 1997), this
critical luminosity may be below the peak outburst luminosity we
observe for SAX J1808.4--3658, however, due to uncertainties in the
distance determinations no conclusive statements can be made.  Also
these critical luminosities appear to be higher in other sources
(e.g. Ford et al. 1997; Ford \& van der Klis 1998; Smale, Zhang, \&
White 1997). However, it is remarkable that when no kHz QPOs were
detected in 4U 0614+09 the break frequency of the band-limited noise
was 0.7 Hz (M\'endez et al. 1997), in the range of the break
frequencies observed in SAX J1808.4--3658. However, when kHz QPOs were
detected in 4U 0614+09 the break frequency was above 6.6 Hz (M\'endez
et al. 1997), which is above our maximum observed break frequency. The
same is true for 4U 1705-44 (Ford et al. 1998). It seems that the
break frequency can be used to predict whether kHz QPOs are observable
or not. If true, we expect kHz QPOs to be observable in SAX
J1808.4--3658 when the break frequency is slightly higher (above
$\sim$7 Hz), thus, when the mass accretion increases to higher values
than we observed during the 1998 outburst. Such higher mass accretion
rates are possible in view of the higher ASM count rates observed
during the 1996 outburst of SAX J1808.4--3658.

We conclude that, although there may be some differences in the
details of how the power spectrum depends on the mass accretion rate,
at this stage there is no evidence for any intrinsic difference in the
rapid aperiodic X-ray variability between SAX J1808.4--3658 and other
LMXBs. So if for example the magnetic field would be considerably
lower in the non-pulsating neutron star LMXBs compared to that of the
millisecond X-ray pulsar, or would be oriented differently, then this
difference in magnetic field does not affect the aperiodic timing
behavior. The conclusion would have to be that the aperiodic X-ray
variability is produced in the accretion disk, and has little
interaction with what happens inside the magnetosphere or even the
magnetopause. If, on the other hand, it is a lower inclination that
allows SAX J1808.4--3658 to pulse where other LMXBs do not, then the
fact that this does not significantly alter the rapid aperiodic X-ray
variability implies a roughly spherically symmetric emission pattern
for these variations, whereas the pulsations would be strongly beamed
along the polar axis. This once again suggests an origin for the rapid
aperiodic variability away from the center, where the pulses
originate, and most likely in the accretion disk outside the
magnetosphere.

\acknowledgments

This work was supported in part by the Netherlands Foundation for
Research in Astronomy (ASTRON) grant 781-76-017. RW and MvK
acknowledge stimulating conversations about the broad-band limited
noise components in LMXBs with Eric Ford, Jeroen Homan, and Peter
Jonker.

\clearpage

\end{document}